\documentclass[epj, nopacs]{svjour}
\usepackage{graphics}
\usepackage{natbib}
\newcommand\Thbb{$T\rm_{HBB }$}

\newcommand{\Msun}{\ensuremath{\, {M}_\odot}}

\newcommand{\Teff}{T$_{\rm eff}$}

\newcommand{\ocen}{$\omega\,{\rm {Cen }}$}
\begin{document}
\title{The pollution from massive AGB stars favoured by strong hot bottom burning}
%\subtitle{Do you have a subtitle?\\ If so, write it here}
%\author{Francesca D'Antona\inst{1} \and Paolo Ventura\inst{1}% etc
\author{Francesca D'Antona \and Paolo Ventura
% \thanks is optional - remove next line if not needed
\thanks{\emph{Present address:} INAF - Osservatorio di Roma, via di Frascati, 00078 Monteporzio (Italy)}%
}                     % Do not remove
%
%\offprints{}          % Insert a name or remove this line
%
%\institute{INAF - Osservatorio Astronomico di Roma \and the second here}
\institute{INAF - Osservatorio Astronomico di Roma }
\date{February 2026}
% The correct dates will be entered by Springer
%
\abstract{Stars of intermediate mass ($\simeq$4-8\,\Msun) evolve to the stage of white dwarfs through the asymptotic giant branch (AGB) stage: stationary hydrogen shell burning and helium thermal pulses, wind mass loss and planetary nebula ejection. Almost the totality of the mass lost (the initial mass minus the remnant white dwarf mass) is heavily processed `hot bottom burning' (HBB), as plain convection reaches the outer edge of the H-burning shell. This phase has been subject of intense investigations in the latest 25 years, in connection to two main research subjects: 1) the chemical evolution of proton-capture elements cycled in these stars, and their intrinsic uncertainties due to the uncertainty in the description of the AGB models;  2) the role of AGBs in the formation of multiple populations in globular clusters. In this work we look back at these problems in the light of the new  possible AGB role in  the composition of hot gas with high N/O in some primordial galaxies, particularly in those hosting a massive black holes.}

%
%\PACS{
%     {PACS-key}{discribing text of that key}   \and
%      {PACS-key}{discribing text of that key}
%     } % end of PACS codes
%} %end of abstract
%
\maketitle
\section{Introduction}
\label{sec1}
In spite of the enormous improvement in the computation of stellar evolutionary models, some of the physical inputs are still not well settled down, in particular there is no theory of mass loss based on first principles, and the overadiabatic convection modeling is mostly still hampered by strong uncertainties, in spite of extended examples of matching with 3D computations. Nevertheless, the `a posteriori' tuning of important parameters on the observations generally are able to provide reasonably accurate matches. \\
The asymptotic giant branch (AGB) stage is the most important site of production of important light elements, but it is among the most difficult phases to be modelled. \\
We will not consider here the AGBs of mass up to $\sim$3\Msun, that evolve into Carbon stars, producing primary Carbon and s-process elements, but will concentrate on the peculiar nucleosynthesis and modelling uncertainties of the upper mass range ($\sim$4-8\Msun). In these stars the convection border enters the hydrogen burning shell, so the whole envelope is progressively subject to proton captures (``hot bottom burning" or HBB), the nuclei resulting from this processing are convected to the stellar surface and lost by stellar winds in the surrounding medium. This elegant scheme was first proposed to explain the luminous Lithium rich giant \citep{cameronfowler1971} discovered in the Magellanic Clouds just above the luminosity limit of C--stars, and is relevant for the chemical evolution of a number of light elements (Nitrogen, Aluminium, Magnesium, and the s-process elements linked to the $^{22}$Ne($\alpha$,n)$^{25}$Mg neutron chain). 
The massive AGBs  were proposed as possible site for the nucleosynthesis of the p-captures which characterize the composition of a large fraction of stars in Globular Clusters (GC), dubbed ``second generation" or 2G stars \citep{ventura2001}. Unfortunately, the structure of AGBs in HBB ---as well as the lower mass range for the case of Carbon stars, for different reasons--- faces the above mentioned problem of a poor understanding of the most important input parameters (Sect. 2).\\
Can we get information to choose the more reliable models among the different possibilities? The ejecta of intermediate mass AGBs can be related with the compositions seen in the 2G stars if 1) HBB is strong, to deplete oxygen ---and in some cases also magnesium---  and 2) mass loss proceeds at high rates, preventing the dramatic increase of the total CNO content due to the effects of the third dredge up, but not found in 2G stars. The models by Ventura et al. \cite{ventura2001}, were the first to obtain these results, as they employed both 1) a very efficient convection model \cite{cgm1996}, consequently reaching higher temperatures at the convective envelope bottom (\Thbb) and 2) a mass loss formulation with a high power dependence on the stellar luminosity \citep{blocker1995}, limiting the total number of stellar pulses and thus the carbon dredge up episodes during the evolution of the models in the upper mass range \citep[see the extensive discussion in][]{ventura2005a,ventura2005b}. We discuss some points of this application of AGB models in Sect.\ref{sec3}.\\
A further opening in this field has been the discovery that the hot gas at the center of high redshift protogalaxies often shows an N/O ratio much larger than expected from the local galaxies of similar metallicity, and not predicted by standard chemical evolution models \citep{marques2024}. This appears to occur especially where the spectra show the presence of an Active Galactic Nucleus (AGN), indicating that the galaxy hosts a massive central black hole (BH) \citep{isobe2025}, and the high N/O ratio is found only in the gas of the central regions, closer to the BH, and not in the general field \citep[e.g.][]{ji2024}. We have shown that this feature can be connected to the formation events in a proto Nuclear Star Cluster: we suggest \citep{dantona2025} that the highest N/O ratios are due to the intermittent super-Eddington accretion phases through which the growth of the central BH has to occur, according to most of the studies following the evolution history of the accreting BH \cite[see, e.g.][]{schneider2023, trinca2023, trinca2024}, but we show why some interesting consequences of this hypothesis require that we can give credit to the CNO abundances emerging from our models (Sect.\ref{sec4}).\\
%In the following we discuss some critical points in this research field, in light of the new challenges posed by high redshift observations.

\section{The widely differing yields of AGBs}
\label{sec2}
\begin{figure}[t]
\centering
\begin{minipage}{0.99\textwidth}
\vskip -33pt
\resizebox{.5\hsize}{!}{\includegraphics{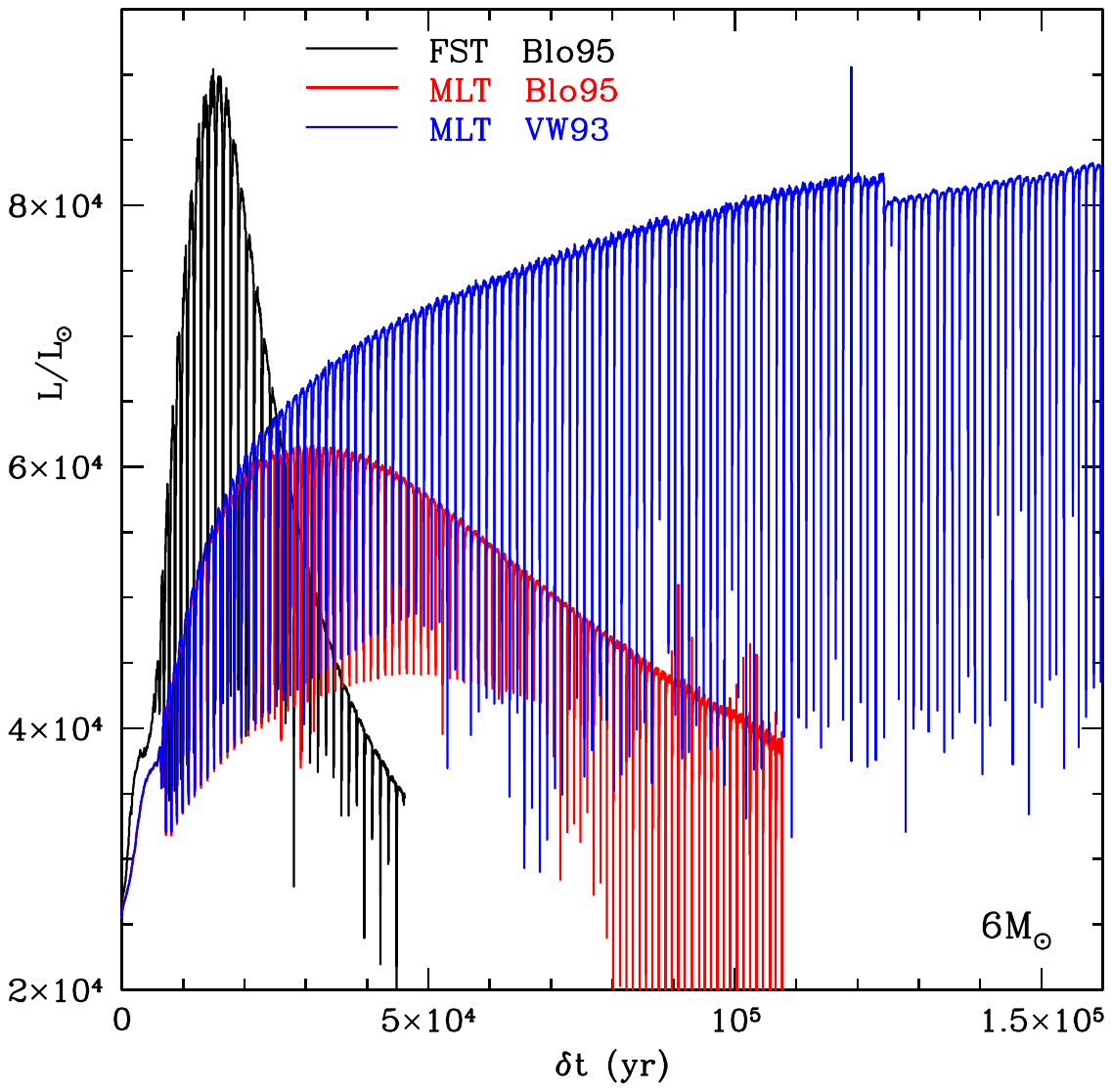}}	  
\end{minipage}
\vskip -55pt
\caption{The dramatically different AGB luminosity evolution of a 6\Msun\ having a metallicity of Z=10$^{-3}$, under three different mass loss modelling \citep[Blo95,][]{blocker1995}, \citep[VW93,][]{vw1993}, and/or convection modelling \citep[FST,][]{cgm1996}, \citep[MLT,][]{bohmvitense1958}. The parameter $\eta$\  in \cite{blocker1995} is fixed to 0.02, and $l/H_p=1.9$\ in the MLT models. The first two tracks are evolved till most of the envelope mass is consumed, while the display of the latter one covers only a small part of the evolution.
}
\label{fig1}
\end{figure}

In this work we mainly concentrate on the uncertainties concerning the evolution of the upper mass range of the stars evolving through the AGB phase. We mention that the stars in the lower initial mass range, those evolving through the Carbon star (C-star) phase, are affected by many other problems which are not touched here, e.g. the role and extent of extra-mixing at the bottom of the convective region, and the efficiency of the `third dredge up', tools necessary to achieve the C-star phase at the correct luminosities. In addition, the C-star phase is associated with a significant increase in envelope opacity, which changes the stellar \Teff\ and the mass loss rate, and finally allows dust formation and envelope ejection. The approach to these problems may affect in part the lower limit of the mass range we are going to examine, the masses subject to HBB.
%internal mixing processes, particularly their influence on the onset and efficiency of the third dredge‑up, as well as the role of surface opacities when the star becomes temporarily carbon‑rich. These factors can significantly affect the temperatures at the base of the convective envelope and the resulting yields.
We will consider together both the masses with degenerate C-O cores up to $\sim$1\Msun, and the ``super--AGB" masses (which undergo an off--center carbon ignition under partially degenerate conditions, followed by a series of thermal pulses, and supported energetically by a CNO burning shell, above an O–Ne degenerate core \citep[e.g.][]{siess2010, doherty2014}) up to the minimum mass exploding as e--capture supernova. In the low metallicity mass range (say [Fe/H] from --2 to --0.7) this mass range is approximately $\sim$4--7\Msun. \\
The key parameter guiding the evolution of massive AGBs and of the super--AGBs is not nuclear burning, but mass loss, generally parametrized as a semiempirical function of the stellar luminosity. But the 
stellar luminosity in the HBB is itself dependent on the description of convection, while the occurrence and importance of HBB affect also the envelope nucleosynthesis and the resulting yields. Thus the main important and badly known parameters we are dealing with are also intertwined each other in a complex way. For this reason, elemental yields from different authors differ significantly each other, even by factors $>$10, in this mass range, with consequences on our understanding of their chemical evolution in different galactic contexts. \\
Here we show this basic problem with new models in Fig.\ref{fig1}, where we compare the temporal luminosity evolution of a 6\Msun\ at Z=10$^{-3}$, computed by assuming either the mixing length theory (MLT) convection scheme with $\alpha=l/H_{\rm p}=1.9$ (the value necessary to reproduce the solar location with present--day opacities) or the ``full spectrum of turbulence" (FST) model by \cite{cm1991, cgm1996}, both employing the \cite{blocker1995} mass loss, and a third track employing MLT and the mass loss by Vassiliadis and Wood \cite{vw1993}. Although the first formulation  \citep{ventura2005a} of this result dates back 20 years, it still remains the main warning signal of why the chemistry of AGB ejecta can differ so much.  \\
\begin{figure*}[t]
\centering
%\includegraphics[width=0.9\textwidth]{fig.eps}
%\begin{minipage}{0.99\textwidth}
\resizebox{.98\hsize}{!}{\includegraphics{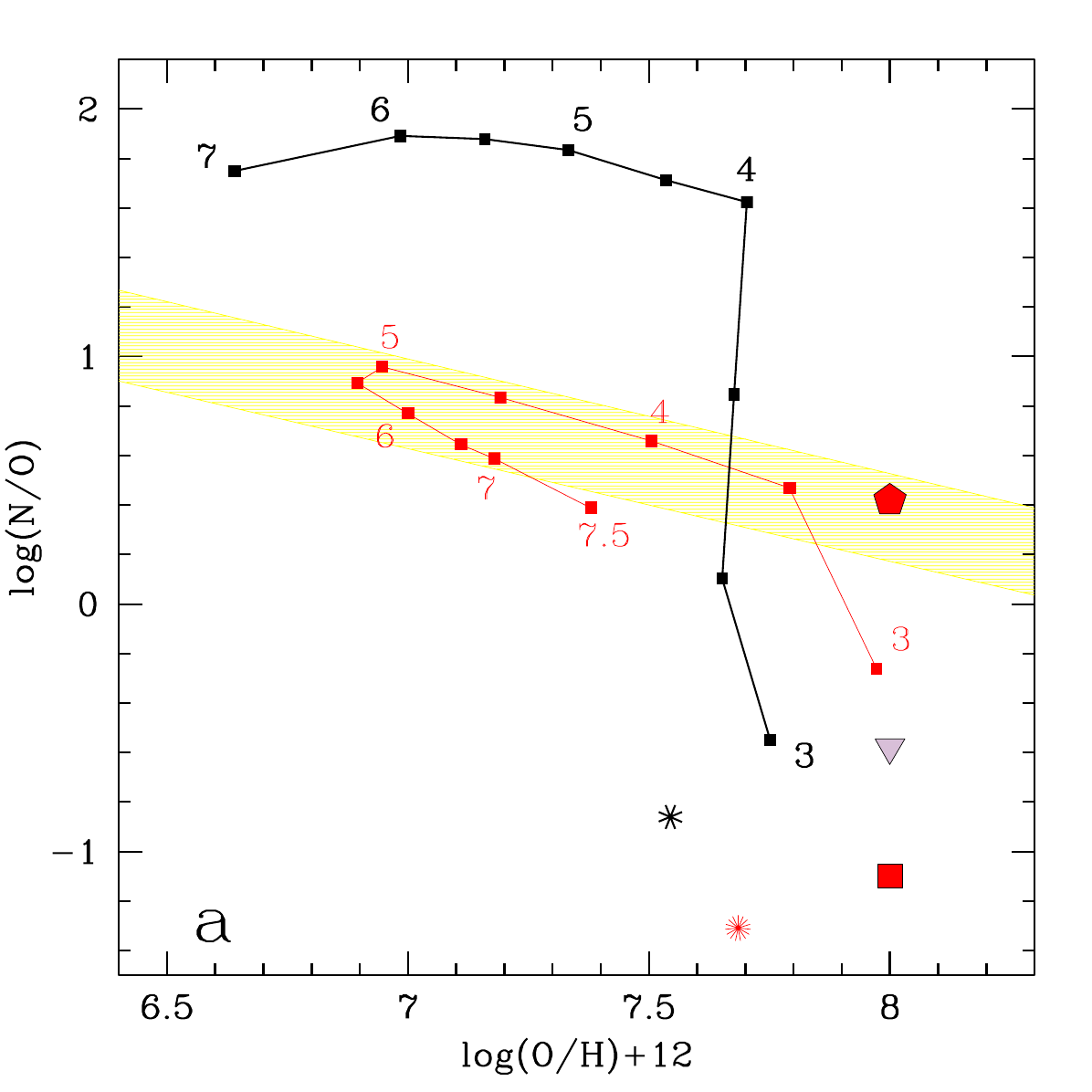} \includegraphics{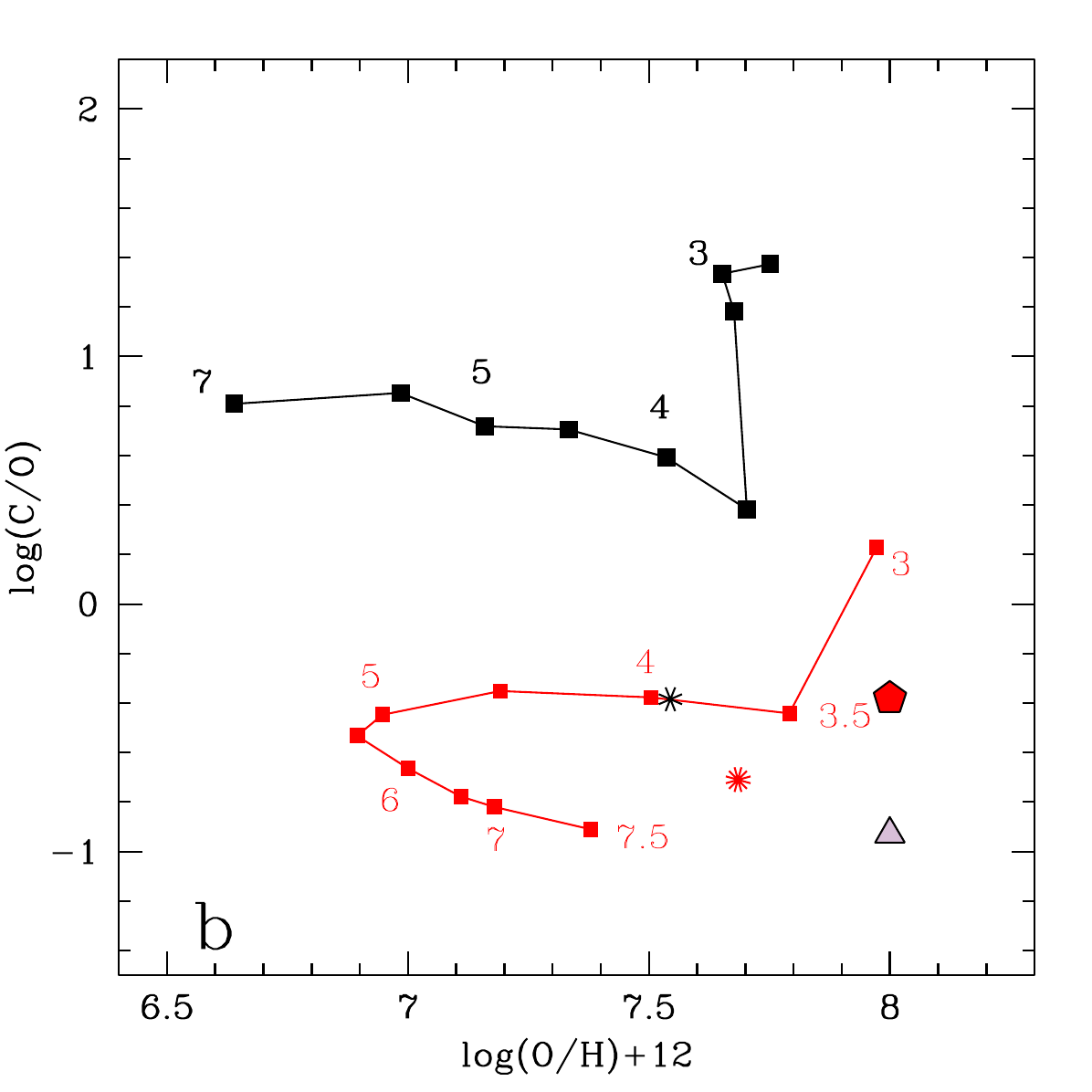}}	  
%\resizebox{.5\hsize}{!}{ \includegraphics{fig-gal1b.pdf}}	 
%\end{minipage}
\caption{We show the average abundances in the ejecta of intermediate mass AGB stars from \cite{ventura2013, ventura2018} (red squares, V13) and from \cite{fishlock2014} (black squares, F14) in the planes log(N/O) (panel a) and log(C/O) (panel b) as a function of log(O/H)+12, a display commonly used to plot abundances, especially from high redshift objects. A few initial masses (in solar units) are labelled on the points, and the abundances refer to a metallicity of Z=10$^{-3}$. The initial O, C/O and N/O abundances (asterisks) for the two sets are a bit different, depending on the specific assumptions of the models concerning the $\alpha$-enhancement and the solar standard abundances, but the resulting ejecta are strikingly different, by more than a factor 10. The yellow strip represents the average location of AGB ejecta of different metallicity in our models, and extends from low to high O/H for increasing metallicities.
The points at log(O/H)+12=8 represent the abundances in GS\_3073 from Ji et al. \citep{ji2024}: the red pentagon refers to the UV lines in the dense region, the inverted grey triangle and the red square refer to the optical region, with different assumptions concerning outflow. Thus the very high N/O, where the pentagon lies, refers only to emission from the higher density part  of GS\_3073,  and is consistent with the composition of pure AGB ejecta of metallicity larger than Z=10$^{-3}$ of the plotted models \citep[see][]{dantona2025}.
}
\label{fig2}
\end{figure*}
Fig.\ref{fig1} helps to understand the comparison of the average composition in the ejecta of models in the whole intermediate mass range (Fig.\ref{fig2}) in the useful planes log(N/O) (panel a) and log(C/O) (panel b) versus the oxygen abundance log(O/H)+12. We show both our model results from \cite{ventura2013} (V13, red squares) and those by \cite{fishlock2014}  (F14, black squares). The metallicity is Z=10$^{-3}$\ for both sets, but some differences in the initial elemental abundances can be seen from the location of the asterisks. Both the choice of a less efficient mass loss formulation (prescription in \cite{vw1993} for F14, in \cite{blocker1995} for V13)
and a less efficient convection model (standard Mixing Length Theory by \cite{bohmvitense1958}, with $\alpha$=l/H$_p$=1.86 for F14, versus FST for V13), with consequently a very long thermal pulse phase (and carbon dredge up) in the F14 models lead to differences of a factor $\sim$10 in N/O and even more in C/O: actually ALL the ejecta from F14 have log(C/O)$>$0, that is they have been Carbon stars for part of the evolution. Important consequences derive from this simple comparison: 1) according to the chosen yields, intermediate mass stars at low metallicity are a source of primary nitrogen and carbon; 2) the amount of s-process elements formed through the operation of the Ne22 chain are also affected by the duration of the AGB phase.\\
A further interesting consequence is that the N/O and C/O resulting from chemical enrichment by AGB ejecta in particular sites, such as the compact primordial galaxies \citep[e.g.][]{rizzuti2025} is very different. This is strikingly clear if we compare the location of ejecta in our models \citep[and their average location for different metallicities, represented by the yellow strip][]{dantona2025} with {\it the location of the gas close to the supermassive BH of the young galaxy GS\_3073, at z=5.5 \citep{ubler2023, ji2024}}. The UV lines coming from the denser regions of  GS\_3073 (probably close to the BH) have abundances lying on the yellow strip, while the optical lines are at much lower N/O values, consistent with a normal galactic evolution. From these data we propose a model to explain the high N/O, based on a temporary emission of gas coming from the AGB wind ejecta only \citep{dantona2025}, during the quiescent phases of intermittent super--Eddington accretion on the AGN. Obviously, this possible scenario is relevant only if our AGB models, but not the F14 models \cite{fishlock2014}, describe the AGB abundances correctly, and it is very important to have experimental checks of which models are more appropriate. 

\section{The AGB models and Globular Cluster formation}\label{sec3}
If we could boldly attribute the formation of 2G stars in GCs to the gas lost from a first generation AGB population evolving in the cluster 30--100\,Myr after the cluster formation, we would gain a powerful way of checking the main physical descriptions in AGB models, and trust the same models for other applications.\\
AGBs as a source of the abundances of 2G stars were proposed already 45 years ago \citep{norris1981, dantona1983}, but the first complete ``AGB model" was systemized in D'Ercole et al. \cite{dercole2008} both concerning the chemical and dynamical predictions.
The dynamical model, based on the onset of a cooling flow when the cluster core collapse supernovae (CCSN) stop exploding, was well confirmed in the following studies \citep[e.g.][]{bekki2011} and in the more recent 3D hydro--simulations \citep[e.g.][]{calura2019} with adequate spatial resolution. The mixing of the stellar populations was also studied in detail, and well matches the observed radial distributions and anisotropies. The composition of the ejecta processed by HBB satisfies a good number of constraints which are unique to this model, and here we summarize those concerning the abundance of light nuclei most relevant for the comparison with different models: 1) the presence of an upper limit to the helium content of 2G stars; 2) the Lithium abundances in 2G; 3) magnesium depletion and its dependence on the metallicity of the cluster stars; 4) how to explain GCs hosting a population of CNO rich stars (Type II clusters). Difficulties of the model are discussed in several works \citep[e.g.][]{renzini2015}.

\subsection{The presence of an upper limit to the helium content of 2G stars}
All possible scenarios for the formation of multiple populations is GCs predict that the stars showing chemical anomalies also must have enhanced helium abundance. In fact, the 2G stars display abundances of light elements (e.g. C, N, O, Ne, Na, Al, Mg) resulting in gas subject to proton captures at high temperatures. This may happen in two main sites: 1) either in the H–burning convective cores of `first generation' (1G) high mass \citep[e.g.][]{decressin2007, demink2009, bastian2013a} or supermassive stars \citep{denissenkov2014} --hereinafter the Convective Core Hydrogen Processing (CCHP) models; 2) or at the basis of the convective mantle of massive AGBs. Helium and the hot-CNO products come to the stellar surface by means of the chemical mixing associated with the transport of angular momentum through the stellar radiative layers in the massive and CCHP stars, and by plain convection in the AGBs, to be lost into the intracluster medium.\\
The main product of H--burning is anyway helium, so also helium must be larger in the processed stellar ejecta. The helium enrichment for the core-H-burning stars thus depends on the stage at which this process occurs, and has in principle no upper limitation, so Y up to $\sim$0.8 is predicted by the massive stars polluters \citep{chantereau2015}. No study is specifically dedicated to the helium content in the ejecta of very massive or supermassive stars, but also in these models there is formally no limitation in the helium content of ejecta which are supposed to participate in the 2G formation.\\
Helium differences among the multiple populations have been observed, in a few specific cases, from the MS location(s): $\simeq$20\% of \ocen\ stars populate a “blue” MS \citep{bedin2004}, soon interpreted as an MS having an abnormally huge helium content \citep[Y$\sim$0.40,][]{norris2004, piotto2005}. In contrast, NGC\,2808 does not show \citep{carretta2006}, or has a very small \citep{legnardi2022} metallicity spread, but displays a similar evidence for the presence of a blue MS, interpreted again as due to a group of stars at Y$\sim$0.40. The MS of this cluster is actually made up of three separate MSs \citep{piotto2007}, the bluest of which, including $\sim$15\% of stars, is consistent with Y$\sim$0.35--0.40. 
In all the other clusters large spreads in Y are not present, as there are nowhere very hot MS stars, corresponding to low mass stars with Y up to $\sim$0.8 \citep{chantereau2016}.\\
%A variation in helium content is immediately reflected in the morphology of the horizontal branch (HB), which amplifies any evolutionary difference among the cluster stars. Helium enrichment by a small factor (up to ) allowed in Y∼ 0.28–0.30 fact an easy interpretation of some puzzles posed by the HB (blue tails, gaps) (D’Antona et al. 2002; D’Antona & Caloi 2004).
The determination of the helium content in the multiple population stars of GCs has known a further development with the use of the ``chromosome maps" \citep{milone2017chromo}, and \cite{milone2018he} compared measurements in the bands of HST photometry to synthetic spectra to constrain the average and maximum helium variations among 1G and 2G populations in 57 GCs. They constrain it to ``more than 0.1" in helium mass fraction (namely the maximum $\delta$Y=0.124$\pm$0.007 is found for NGC\,2808). In other words, there is no evidence, in any cluster, of stars having a helium abundance unlimited as it could be expected by the ejecta of massive stars. Consistency with this observational result is anyway the model of accreting Extremely Massive Stars  \citep[aEMS,][]{gieles2025}. They establish that the helium in their 2G results from the ejecta of very massive (VMS) and extremely massive (EMS) stars, increasing with time up to Y=1, diluted with the accreting pristine gas having standard Y. The models they display have a low average $\delta$Y$\sim$0.09, and a maximum $\delta$Y$\sim$0.12, as observed. We should be aware that this result can vary with the accretion timing and with the distribution of the IMF. 
Even more suspicious is the maximum very similar Y of the populations spanning $\sim$1.5\,dex in \ocen\ \citep{clontz2025}: for all of them should it be a result of very similar IMF extending to EMS?
\\
The case of AGBs ejecta is indeed very different, because the helium enrichment is mainly due to the second dredge-up (2DU), which precedes the thermal pulse phase \citep{ventura2010}. Also, the third dredge-up (3DU) episodes during the thermal pulse phase could contribute to helium enrichment in the envelopes, but a further observational constraint is that the sum of CNO abundances is remarkably the same (at least, within a factor $\sim$2) in normal
and anomalous stars \citep[e.g.][]{cohenmelendez2005}. Consequently, the masses which can be responsible for the chemical enrichment, and their mass loss rates,  must be high enough that they suffer only a few episodes of the 3DU, which brings into the envelope the products of helium burning. 
The values of Y in the massive AGB ejecta are a bit different in the results of different authors \citep[e.g.][]{siess2010, ventura2013, doherty2014}, nevertheless they have a very special common feature: for the whole range of massive AGBs and super-AGB stars, the average helium in the ejecta attains a maximum value in the range Y=0.35--0.38, also for the masses which have the most extreme p--processing of the envelope. These values are close to those inferred by the above interpretation of the observations of extreme populations. \\
We conclude that the presence of an upper limit for the observed helium abundance in 2G stars, close to the upper limit of the helium content of massive AGB and super-AGB ejecta\footnote{Needless to say, the uncertainties both in the determination of helium abundances and the in models do not allow to take seriously a difference of a few 0.01 in their values as noted in \cite{doherty2014}, fig.8a.} is one of the most appealing feature of the AGB model for the formation of multiple generations in GCs.

\subsection{Lithium abundances in 2G}
Apart from the AGB scenario, in all the models proposed for the formation of 2G stars, the p-processed gas comes from the stellar interior; thus this gas has totally burned its primordial lithium. On the other hand, in the AGB model, the same HBB which is responsible for the anomalous composition of the 2G stars, also produces fresh lithium by the Cameron and Fowler \cite{cameronfowler1971} mechanism. Li remains in very high abundances in the envelope until $^3$He is completely consumed, and eventually it is fully burned. Thus, the massive AGBs can expel Li with the gas lost during their initial HBB phase and contribute to the Li abundance of the mixture.  A problem with this hypothesis is that the Li average abundance in the ejecta is scarcely constrained, due to its strong dependence on the mass-loss rate during the lithium rich phase \citep{ventura2005b, dantona2019}. Thus, while in fact the presence of lithium in 2G stars is often regarded as a good hint in favor of the AGB model, dilution of the polluter matter at zero Lithium can achieve similar results and the argument is not conclusive \citep{dantona2012, dantona2019}. Different is where we find that giants with extreme anomalies (and, possibly, also independently derived helium content at the top of the values found in 2Gs) have a high lithium content, such as the ``extreme'' giant ID$\#$46518 \citep{dorazi2015}, belongs to the `extreme' group in NGC\,2808 and has $\log \epsilon$(Li)=0.96$\pm$0.1. \cite{dantona2019} analized this result and show that the abundances in this giant are compatible with its direct birth from super--AGB ejecta, while its Li abundance requires a large degree of dilution with pristine, Li-rich, gas, if it was born in the CCHP scenario.

\subsection{Magnesium depletion and its dependence on the metallicity of the cluster stars}

%\citet{ventura2018} summarize the problems of proton captures on Mg isotopes in HBB AGBs. Problems are met with the isotopic ratios observed in the giants of 1G and 2G stars in the cluster NGC\,6752, and they can be tentatively solved when the p--capture rate by $^{25}$Mg at the $\sim$100\,MK, and in particular the $^{25}$Mg(p,$\gamma$)$^{26}{\rm Al}_{\rm m}$ channel, is enhanced by a factor $\sim$3 with respect to experimental determinations.\\
A clear correlation holding for the AGB model is confirmed by the comparison with the Mg-Al anticorrelation extent in GSc from the Apogee survey \citep{ventura2016}: Mg depletion in the 2G is expected only for low metallicity GCs, and especially for GCs hosting extreme 2G populations in general (e.g. population with helium at its maximum values) \citep{ventura2018}. This is due to two simple reasons: first, Mg burning requires \Thbb$>$100\,MK\footnote{Central temperatures above $\sim$80\,MK are needed to deplete Mg in the H--burning cores} and \Thbb\ depends on the convective envelope opacities, so it is larger for lower metallicity. Second, dilution with gas with standard composition reduces the Mg depletion. \\
Possibly the best example of this behavior is found in \ocen, a former Nuclear Star Cluster where metallicity spans more than a decade. Alvarez-Garay et al. \cite{alvarezgaray2024} showed not only that Mg depletion increases for decreasing metallicity of the samples examined, but that at the same time Si abundance increases, and Al is smaller, testifying the extension of the p--burning chain to the reaction $^{27}$ Al(p,$\gamma$) $^{28}$Si at low metallicity.  
\\
Magnesium depletion is not predicted by CCHP models, apart from the supermassive stars \citep{denissenkov2014}, which were proposed to account for it.  The aEMS \citep{gieles2025} model extends the clusters' IMF to the mass range of accreting extremely massive stars (EMS) of 10$^3$--10$^4$\Msun. In clusters of mass $>$10$^5$\Msun, this mass range contributes to the formation of the 2G. In the aEMS models, Mg depletion is expected for those clusters initially very massive (where EMS of $>$3000\Msun\ are expected) and of low metallicity, as in fact found. The comparison with GC data shows a fair agreement, so we can consider this an alternative scenario for the 2G. The examples studied in Gieles et al. \cite{gieles2025} do not touch the ingent Mg depletions of at least 1\,dex in NGC\,2419 \citep{cohenkirby2012}, a massive cluster with very low metallicity ([Fe/H]=-2.1), nor the anticorrelation Mg--Si, for which the Al abundance is not necessarily at its maximum values when Mg is mostly depleted and well explained in the AGB model \citep[see, e.g.][]{ventura2018,alvarezgaray2024}.

\subsection{The problem of GCs hosting a population of CNO rich stars (Type II clusters)}
A final peculiarity of multiple populations is posed by the presence of a generally small 2G with C+N+O and s-process abundances higher than the value in the the 1G and in the standard 2G. 
%%%%%%% COPIATO DA BEDCA
As we have mentioned, already at early times it was recognized that the total C+N+O abundance is approximately constant in the individual clusters examined, and that the C, N and O variations are due to the action of the CNO cycle, and in particular of the ON branch, which accounts for the oxygen reduction in the 2G stars. 
%Examples of such clusters are M92: \cite{pilac1988}; NGC~288 and NGC~362: \cite{dickens1991};  M~3 and M~13: \cite{smith1996}; M4: \cite{ivans1999}; NGC~6752 \citep{yong2015}. Also \ocen\ shows an increase in s--process and CNO abundances \citep{marino2012cno}, but its evolution may have been more complex anyway. \\ 
%%%%%%%%%%%%%%% anche questo va eliminato?
%A more complex situation was appreciated more recently: first, the splitting found in the subgiant branch of the cluster NGC\,1851 \citep{milone2008} was easily justified by assuming that its 2G (populating the dimmer branch) had a larger C+N+O than the bright branch, but a similar age \citep{cassisi2008,  ventura18512009}. Several other clusters have been shown to hold a double subgiant branch \citep{piotto2012}. The spectroscopic enrichment in total CNO was established in NGC~1851 \citep{yong2009, yong2015},   and M22 \citep{marino2012m22}, see also \cite{lim2015b}. In clusters showing a split sub giant branch, the feature is accompanied by an increase in s-process elements abundance.\\
%, which is in line with the 3DU interpretation \citep{straniero2014}. 
%%%%%%%%%%%%%%%%%%%%%%%%%%%%
%So we have to understand the formation of 
There are other multiple populations, collected  in a category of clusters named ``s--Fe--anomalous"  \citep{marino2015}, and later on also named ``Type II clusters" for their  typical behaviour in the ``chromosome maps". In these clusters, the typical signature of p--capture elements variations (the O--Na anticorrelation) is accompanied by metal enrichment, CNO  and s--process enrichment together. These are mainly clusters with split sub giant branch, namely NGC~1851 \citep{carretta2011, gratton2012sgbs1851}, NGC~5286 \citep{marino2015}, M~22 \citep{marino2009} and M2 \citep{yong2014}, plus \ocen (see later). \\
%Further investigations are needed, to explore the processes which may explain why some clusters show two main bursts of SF, separated by a few tens of million years to allow the CNO--, s--enhanced ejecta production. Nevertheless, 
Production of s-process elements and enrichment in carbon point to nucleosynthesis occurring on long timescales, so explaining these populations in the framework of 2G formation in the short lifetime of the evolution of EMS is not straightforward \citep{gieles2025}.
On the contrary, it may find a coherent explanation in the AGB model, for clusters where star formation is shortly delayed by late supernova explosions, so that the final burst of star formation occurs in gas enriched by AGB winds enriched in s-process and Carbon by the third dredge up \citep{dantona2016}.
%Note also that in these clusters the O--Na anticorrelation is already present in the fraction of stars having the lower, and homogeneous, iron content. In the AGB model, this anticorrelation is very unlikely to occur, if the iron contamination is due to the contribution of SN\,II ejecta. So the first phases of evolution are similar to what happens in GCs which are fully chemically homogeneous in heavy elements, and the ``standard" 2G was formed during the evolution of the massive AGBs.  At later times, we must account for contamination by further supernova explosions, either from delayed SN\,II in binaries or by the first SN\,Ia explosions. Afterward, delayed SN\,II begin exploding with some regularity in a cluster, destroy the cooling flow, which would have included both AGB ejecta and pristine gas, but these events are much less frequent than the single SN\,II and are not able to inject into the gas enough power to definitely push it out of the cluster vicinity. 
When the delayed events become rare, possibly several tens of Myr later, the pristine plus AGB gas will re-accrete and induce a new 2G formation burst. In these hypotheses, {\it the pristine gas will now be contaminated by the delayed SN\,II ejecta, and by the AGB ejecta which were lost during this time span}. The contaminating AGBs will be then the masses in which the 3DU has been very effective, and this new population will have the characteristics of the ``s--Fe--anomalous" clusters: larger iron and s--process abundances, and associated C+N+O enhancement.
An appealing feature of this scenario is that it explains the features of the anomalous clusters, without the need for additional hypotheses, such as the merging of two  different clusters. 
Analytic dynamical models have confirmed this possibility \citep{dercole2016}.\\
A time gap of a few 10$^7$yr  between the formation of the `first', standard 2G, and the `second' one, s-Fe and CNO enriched, also justifies another important characteristics of some clusters: the presence of separate subgiant branches. 
While this time break is negligible in terms of location of isochrones with identical chemical composition, this short time is sufficient to shift the AGB ejecta composition to the CNO enriched stage, which, also with the help of the small iron increase, will result in distinct subgiant branches \citep[for the case of NGC~1851, see][]{cassisi2008, ventura18512009}.
Finally notice that the formation epoch of this s-Fe-CNO enriched 2G can not be very extended, as it occurs close to the beginning of the SN\,Ia era, which will definitely end star formation. 
\\
We must at least remember that \ocen\ might be the extreme case of these GC with extended anomalies, but its internal [Fe/H] abundance dispersion is not limited to a fraction of stars, and does not concern a tiny difference, as in NGC\,1851, but extends for about 2\,dex \citep{johnsonpilac2010,marino2011wcen}. 
As suggested by \cite{dacosta2015}, these systems might be the former nuclear star clusters of now disrupted dwarf galaxies \citep{willmanstrader2012, marino2015}. An iron difference among cluster stars implies that the cluster was able to retain at least some of the supernova ejecta. This may occur either in dwarf galaxies, where possibly dark matter is initially present, or in particularly massive clusters, and it may be helped by the presence of a central massive or supermassive BH.\\
The peculiarity of \ocen\ stars is that {\it for each metallicity} multiple populations are present, also showing very interesting patterns. 

\section{Signatures of GC formation in the high--z galaxies? } \label{sec4}
In the latest years, {\it James Webb Space Telescope} (JWST) observations are revealing a growing population of galaxies at redshift z$>$4 with elevated nitrogen-to-oxygen ratios detected in the emission lines. In the `N/O-enhanced' (NOEGs) galaxies, the ratio N/O is supersolar at subsolar O/H, so it does not conform to the observed (N/O) versus (O/H) abundance pattern of star-forming galaxies in the Local Universe, compatible with a chemical evolution due first to massive stars and later on to the secondary and primary N produced by intermediate mass stars \citep{vincenzo2016}. These peculiar N/O abundances resemble Globular Cluster abundances as has been remarked in many early works \citep{senchyna2024, charbonnel2023, dantona2023}. Ji et al. \cite{ji2026} showed that indeed the similarities between the two populations include the abundances of C, N, O, Fe and He, and are compatible with the scenario in which globular cluster stars formed within proto--cluster environments similar to those traced by NOEGs, that were self-enriched.\\
These new observations offer two points of view: 1) indeed we may have found the 2G formation in the act, and the model(s) explaining the high N/O in NOEGs apply also to the bulk of GCs with multiple populations; 2) we are in the presence of a different kind of event in the dense region of forming galaxies, only in peculiar cases the high N/O signals the formation of stars which later on will result as 2G in a cluster, while in many other cases the rise in N/O is only temporary.\\
In fact, the high N/O is not the most typical feature of 2G, as it may happen at any time we look at gas processed in stellar interiors through the NO branch of the CNO cycle, which occurs at temperatures well below those necessary to deal with the p-captures signs of the 2G stars, such as Mg depletion.
In fact, some of the scenarios proposed to explain the NOEGs gas are not able to provide a consistent 2G models. For example, the suggested of intense and densely clustered star formation with  a high concentration of Wolf-Rayet stars \citep{senchyna2024}, in a single or double stellar burst model \citep{kobayashi2024} represent a short phase in the life of massive star, while even the model of fast massive rotating stars, in which the phase of high N/O mass loss is long lasted \citep{decressin2007} is no longer regarded as a viable 2G formation model.
 Also the proposal of stellar collisions or tidal disruptions in a dense cluster \citep{cameron2023} is not a model for 2G formation. There remains the role of
very massive stars of 100--400\Msun \citep{vink2023}, supermassive stars (M=10$^3$--10$4$\Msun) \citep{charbonnel2023,nagele-umeda2023}, possibly of  Population III   \citep{nandal2024, nandal2025}.
The role of AGB winds has been proposed by D'Antona et al. \cite{dantona2023} in connection to the possible formation of a Nuclear Star Cluster, whose dynamical role is to retain the stellar black holes born in the first phases, merge them and let them grow until they will be supermassive black hole, more than a typical GC. Later on it has been confirmed that many of the NOEGs show signatures of active galactic nuclei (AGNs) \citep{isobe2025}, suggesting a possible connection between black hole formation and nitrogen enrichment.
%watanabe2024, kobayashi2024, dantona2023, maio2024gnz11,
The connection between NOEGs, the formation of NSCs and the role of AGB winds has been confirmed by D'Antona et al. \cite{dantona2025}, by examining the extreme N/O in GS\_3073 \citep{ji2024}, and noting that it is typical of the extreme abundances in the winds of massive AGBs for the (O/H) assigned to GS\_3073. Indeed this extreme abundance is straightforwardly explained by {\it pure AGB ejecta}, leading to suggest a model in which the central AGN accretes in super-Eddington bursts \cite{madau2014}, sweeping out the closeby region, which is then replenished first by the winds of the nearby AGB stars of the NSC. It is precisely in the dense region close to the central BH that the N/O is very high (red pentagon in Fig.\ref{fig2}, while it is about normal in the galaxy field (red square). This event will occur several times during the BH accretion route, at each cycle of super-Eddington accretion. For example, \cite{madau2014} show evolution histories with three main episodes of accretion, at rates three times the Eddington rate, lasting 50 Myr and followed by quiescent periods of 100 Myr, while \cite{volonteri2015} argue for duty cycles of intermittent phases of super-Eddington growth, with accretion episodes lasting 10$^3$--10$^5$\,yr, and `flow regeneration periods' of 10$^4$-10$^5$\,yr.\\
This simple model is confirmed by elaborated THESAN-ZOOM simulations \citep{mcclymont2026} showing that bursty star formation naturally generates order-of-magnitude excursions in N/O on $\leq$100\,Myr timescales due to differential galactic winds; after a starburst, stellar feedback expels gas, leaving a large population of asymptotic-giant-branch stars to dominate the enrichment of the relatively low-mass interstellar medium.
Note that the model by \citep{mcclymont2026} shows that the carbon abundance trends {\it are not correctly reproduced in their models}: this may be due to the use of AGB yields with strong 3rd dredge up, as we have seen in Sect.\ref{sec2}, and testifies the need to use models tested in comparisons with  other environments. 
\\
So, we think that the high N/O in the NOEGs is not a straightforward signature of 2G formation in GCs, and in particular

 1) it is due to high N gas of massive stars, e.g. Wolf Rayet, or other, in the bursting young galaxies in which a high star formation rate is evident from the analysis of the spectrum;

 2) it is more probably due to AGB winds in NOEGs hosting an AGN \citep{isobe2025}, where we must be witnessing a relatively long phase in the galaxy life, necessary to accrete the original black hole seed to the highest AGN masses;
 
 3) star formation in these latter objects, anyway, will not end with typical GC multiple populations, but with more complex abundances patterns, including large variations of iron abundances, more similar to the atypical case of \ocen\ \citep{dantona2025}.\\
Here we have discussed that the interpretation of these trends depends on the yields provided by the widely different computations of AGB models, and that yields consistent with the AGB model for the formation of the 2G in globular clusters give a stimulating insight into the problem of accretion on supermassive BH.
\\ \\
{\bf Acknowledgements}\\
This work is dedicated to the memory of Roberto Gallino, a lucid mind, acute and curious researcher, and great man and colleague.

% BibTeX users please use
%\bibliographystyle{unsrt}
\bibliographystyle{plain}
\bibliography{gallinorev}
\end{document}